\documentclass[useAMS,usenatbib,usegraphicx]{mn2e}
\usepackage{color}
\usepackage{hyperref}
\usepackage{footnote}

\title[X-ray thermal emission in cluster winds with a SN]
 {The soft and hard X-rays thermal emission from star cluster winds
   with a supernova explosion} 
\author[Castellanos-Ram\'irez et al.]
{A.~Castellanos-Ram\'irez,$^1$\thanks{antonio.castellanos@nucleares.unam.mx} 
A.~Rodr\'iguez-Gonz\'alez,$^1$ A.~Esquivel,$^1$
\newauthor{ J.C.~Toledo-Roy$^1$,  J.~ Olivares,$^2$ \& P.F.~Vel\'azquez$^1$}\\
$^1$Instituto de Ciencias Nucleares, Universidad
Nacional Aut\'onoma de M\'exico, Apdo. Postal 70-543,
04510, M\'exico, D.F., M\'exico\\
$^2$Univ. Grenoble Alpes, IPAG, F-38000 Grenoble, France
CNRS, IPAG, F-38000 Grenoble, France}

\begin{document}
\date{Draft Version, \today}
\pagerange{\pageref{firstpage}--\pageref{lastpage}} \pubyear{2007}
\maketitle
\label{firstpage}
\begin{abstract}
Massive young star clusters contain dozens or hundreds of massive stars
that inject mechanical energy in the form of winds and supernova
explosions, producing an outflow which expands into their surrounding
medium, shocking it and forming structures called superbubbles.
The regions of shocked material can have temperatures in excess of
10$^6$ K, and emit mainly in thermal X-rays (soft and hard).
This X-ray emission is strongly affected by the action of thermal conduction,
as well as by the metallicity of the material injected by the massive stars.
We present three-dimensional numerical simulations exploring these two effects,
metallicity of the stellar winds and supernova explosions, as well as
thermal conduction.

\end{abstract}

\begin{keywords}
Hydrodynamics -- galaxies: star clusters: general -- (Galaxy:) open clusters and associations: general -- ISM: bubbles -- shock waves -- stars: massive -- stars: winds, outflows -- X-rays: ISM
\end{keywords}
\section{Introduction}

It is well known that masive O-B type stars inject a considerable
amount of mechanical energy into the interstellar medium (ISM), in form
of stellar winds or supernova (SN) explosions. The energy input by
these events is sufficient to drive strong shocks that expand
into the ISM generating a structure called bubble.

The model proposed by \citet{Weaver1977} and later expanded by
\citet{Chu1990} and \citet{Chu1995},  is considered the standard model of
bubbles driven by stellar winds. It considers the injection
of mechanical energy to the ISM from stellar winds that results in the
formation of a bubble. This bubble is surrounded by a cool shell
of ISM material that has been swept by the expanding shock front.
The shocked (and thereby heated and compressed) material in the interior of the 
bubble emits considerably in X-rays, whereas the outer, cooler shell
emits at optical wavelenghts. 

The original \citet{Weaver1977} model considers a single stellar wind
source. Some time later, in order to explain what is now known as
  \emph{superbubbles}, the model was extended to include multiple wind sources
  \citep[see][]{Chu1990,Chu1995,Canto2000,Silich2004} .

The simplest model of superbubble formation is as follows. Consider a cluster with $N$
stars each having different mass-loss rate $\dot{M}_{\mathrm{w},i}$ and a
wind velocity $v_{\mathrm{w},i}$. Since the stars inject mechanical energy in the form of
stellar winds, the total mechanical luminosity is given by 

\[
L_\mathrm{w}=\sum^{N}_{i=1}\frac{1}{2}\dot{M}_{\mathrm{w},i}v^2_{\mathrm{w},i}\,.
\]

At first, the stellar winds collide with each other and with the
enviromnent inside the cluster radius. Thus the space
between the stars is filled with hot shocked material from the winds.
This happens until a stationary flow is established,
giving rise to a common cluster wind that forms a
\emph{supershell}. As this supershell expands through the surrounding
ISM it creates a superbubble structure with the following structure
\citep{Weaver1977,Rodriguez-Gonzalez2011,Velazquez2013}:  

\begin{enumerate}
\item The innermost region located near the stars (where their winds
  collide) produces thermal hard X-ray emission (if the stellar winds
  have terminal velocities larger than $1000$ km~s$^{-1}$), and driving the
  expansion of the bubble through a pressure difference between the hot and
  dense interior and the colder and less dense environment.

\item After the individual winds from the stars coalesce into a
  cluster wind, it expands freely from the cluster radius outwards. In
  this zone X-ray emission is important only close to the cluster
  radius, and it consists mostly of soft X-rays.

\item Behind the main shock pushing into the ISM a reverse shock is
  formed. The reverse shock encounters the freely expanding wind and compress and  heats it to soft X-ray emitting temperatures. The region filled by
  shocked wind is quite extended and dominates the emission in X-rays,
  particularly in the soft energy bands.

\item The outermost region of the supperbubble consists of a shell of
  shocked ISM that has been swept up by the main
  shock. Beyond this zone there is only unperturbed ISM material.

\end{enumerate}

The original wind blown bubble (WWB) model proposed by Weaver et al.
(1977) overpredicts the X-ray luminosity. One reason is that this model
includes thermal conduction and in consequence produces a denser interior
that in the case without it, which in turn increases the X-ray luminosity.
Furthermore, the wind blown bubble models do not take into account the
radiative losses within the cluster radius, and this can have a
significant impact on the luminosity (see Rodr\'iguez Gonz\'alez et al.
2011). Recent examples of this are the works of
Dunne et al. (2003) and Reyes-Iturbide et al. (2009), which predict X-ray
luminosities that exceed that of the observations by about one order of
magnitude.

On the other hand, there are others models that predict an X-ray emission
that underestimates the observed values (for instance, see the work of
Harper-Clark \& Murray 2009; Rogers \& Pittard 2014, in which only the
cluster wind region is considered), a problem for which different
solutions have been explored. \citet{Chu1990} proposed that in order to increase the luminosity of
X-rays (so as to match the observations) one should consider
shock waves produced by the explosion of supernovae inside the star
cluster. 
\citet{Stevens2003} presented models where the luminosity of soft
X-rays is obtained as a function of the mass loss rate, the cluster
radius and the wind terminal velocity.
They do not take into account mass loading, but they
consider it can be relevant for the study of soft X-rays in this type
of massive clusters. The work of \citet{Silich2001}  deals with the
effect on the X-ray emission of the high metal content injected by the massive stellar
winds and the SN remnants.
\citet{Rodriguez-Gonzalez2011} showed that  supernovae occurring 
near the centre of  the cluster are not capable of reproducing
(completely) the luminosity observed in X-rays, and neither do they help
explain the kinematics of the shell (without consider thermal conduction). They instead showed that
off-centre SN explosions (for N70 and N185, see also
\citealt{Reyes2014}) could help explain the two or three orders of 
magnitude difference between the luminosity observed and the standard
model predictions. However, in these models the X-ray 
luminosity agrees with the observed value for only $\sim 10\,000$
years, making the probability of observing them in this regime rather low.
On the other hand, \citet{Velazquez2013} presented models of the  M\,17 superbubble where they
considered the contribution of the gas of the parental cloud in the
evolution. They showed that the mass
loading from the parental cloud can help increase the luminosity
of soft X-rays by up to an order of magnitude. In such work neither the
metallicity nor thermal conduction were considered.

The Large Magellanic Cloud (LMC) is filled with superbubbles
with important soft X-rays emission. Some of these superbubbles
(for instance, DEM L50 and DEM L152, see
\citealt{Jaskot2011}) show evidence for off-centre supernova events
which seem to interact with the external shell pushed by the stellar
winds. In the observations made by \citet{Jaskot2011} the gas of  the
supernova remnant is still seen. This remnant is located close to the
superbubble edge. These objects have luminosities up to an order of magnitude
higher than those predicted by the model of \citet{Weaver1977}.

Moreover, the numerical models that  appear in \citet{Jaskot2011}
produce luminosities that are two orders of magnitude lower than
the observations ($\sim10^{36}$ erg s$^{-1}$). Therefore the authors
explored the effect of the metallicity and of mass loading by
clouds to bridge the luminosity deficit in soft X-rays.
They calculated the mass of metal enriched material injected by the
supernova explosions \citep{Maeder1992, Oey1995, Silich2001, Anorve2009} and found that
metallicities from 3 to 10 times solar can be achieved, and using the
equations of \citet{Silich2001}, they concluded that the effect is not
sufficient to account for the differences. They conclude that the main
mechanism that can explain such an important enhancement of the total
X-rays luminosity  is mass loading.

Recently, \citet{Rogers2014} presented a study of the soft
X-ray emission during the various evolutionary stages of massive stars
embedded in a dense giant molecular cloud (GMC), going through the red
supergiant and Wolf-Rayet stages up to the supernova phase. They
showed that the inclusion of the GMC results in a short lived
attenuation of the X-ray emission of the cluster, during the time
before an important fraction of the material is carried away from
the wind interaction region. After this occurs, the luminosity 
remains practically constant.
 
The X-ray emission of a star changes substantially as it goes through distinct
evolutionary stages. For instance, the X-ray luminosity drops
abruptly during the red giant phase and increases substantially once in the Wolf-Rayet
phase. \citet{Rogers2014} show that, in spite of the differences between their
models and some observations, their results agree reasonably with other
observations, such as the case of M17 and the Rosette Nebula.
They found that the emission produced by their model during
the early wind-dominated phase is smaller compared to the prediction
from the standard model  \citep{Weaver1977, Chu1990}, but larger than
the emission expected in  models that only consider the emission at
the interaction region of the winds of massive stars (the cluster wind). 
Finally, for stars in the main sequence, they found luminosities two or three
orders of magnitude above those predicted by the standard model,
lasting for more than $4.5~\mathrm{kyr}$.

In this work, we present a series of numerical models
exploring the effects of supernova explosions, metallicty and
heat conduction in the thermal soft and hard X-rays luminosity of a
massive star cluster. The paper is organised as follows: in Section 2
we present the numerical setup of our models,
describe the implementation of the thermal conduction and the
metallicity in the gas dynamics equations, and 
in Section 3 we show the resulting synthetic emission in the soft and hard X-ray
bands as well as a brief discussion of our results . In Section 4 we have made some comparisons of our numerical models with four interesting observed bubbles. Finally, a summary is given in Section 5.

\section{The numerical simulations}

With the purpose of exploring the effects of the interaction and influence
of the SN explosions and metallicity, as well as thermal conduction
in the cluster stellar winds, we performed a series of numerical
simulations, and estimated the soft and hard X-ray emission that would
be produced. 

\begin{figure}
\centering
\includegraphics[width=\columnwidth]{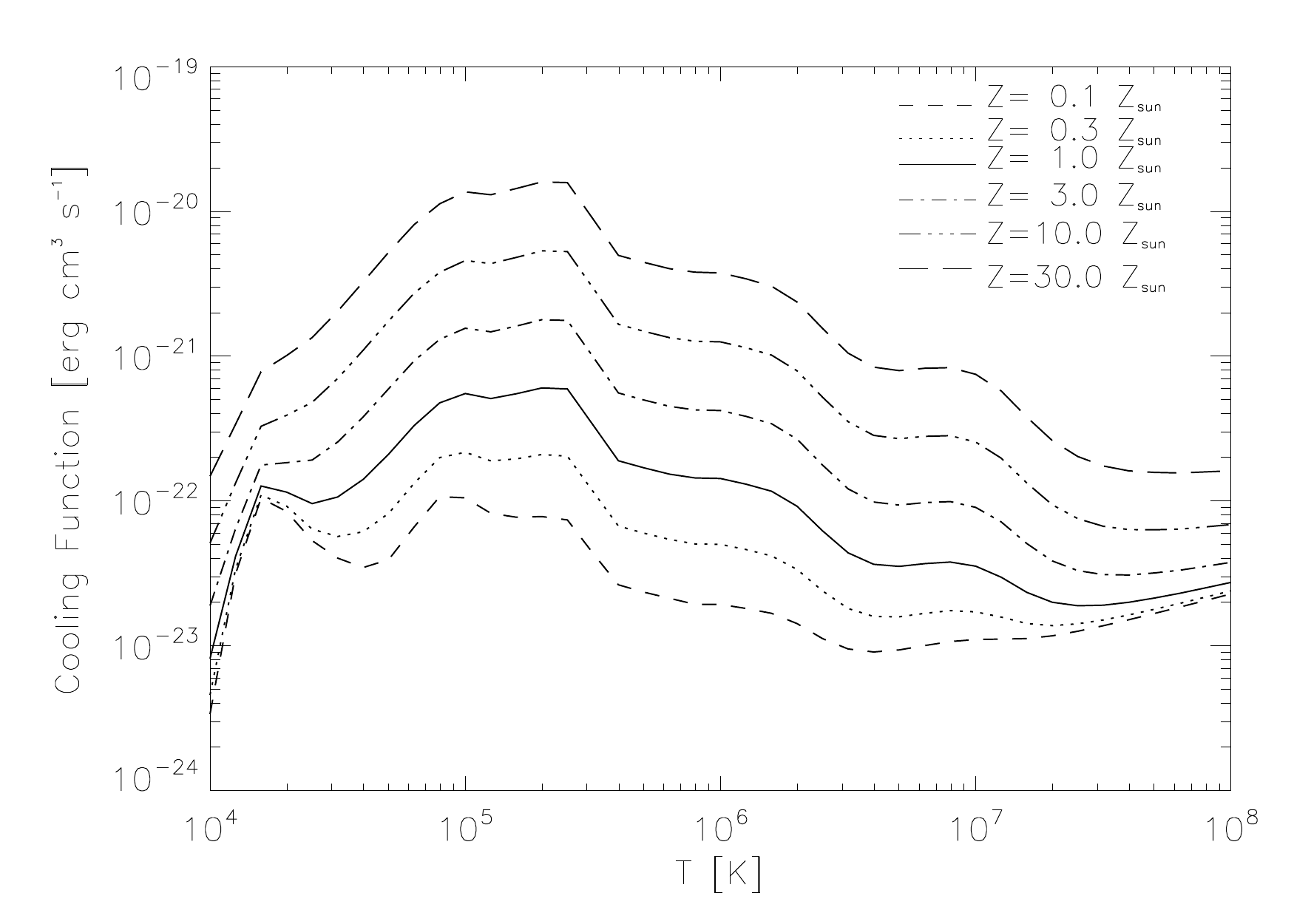}
\caption{Cooling function for a range of metallicities between 0.1 to 
30~Z$_{\odot}$.}
\label{f:cooling}
\end{figure}

We used the {\sc huacho} code (see \citealt{Esquivel2009} \& \citealt{Raga2009}) to
perform all numerical simulations. The code solves the hydrodynamic
equations (\ref{eq:conmass}-\ref{eq:conene}) on a three dimensional
uniform Cartesian mesh, using a second order finite volume
  method with HLLC fluxes (\citealt{Toro1994}) and a piecewise linear
  reconstruction of the variables at the cell interfaces with a minmod
  slope limiter. The code also includes radiative losses and
isotropic thermal conduction: 
\begin{equation}
\label{eq:conmass}
\frac{\partial \rho}{\partial t}+\mathbf{\nabla}\cdot(\rho \mathbf{u})=0,
\end{equation}

\begin{equation}
\label{eq:conmom}
\frac{\partial \left( \rho \mathbf{u}\right) }{\partial t}+\mathbf{\nabla}\cdot
\left(\rho \mathbf{u}\mathbf{u}+ \mathbf{I}P \right)= 0,
\end{equation}
\begin{equation}
\label{eq:conene}
\frac{\partial E}{\partial t}+\mathbf{\nabla}\cdot
\left[ \mathbf{u}\left(E+P\right)\right]=
L_\mathrm{rad}(Z,T)+\mathbf{\nabla}\cdot \mathbf{q},
\end{equation}
where $\rho$, $\mathbf{u}$, $T$, $P$ and $E$ are the mass density,
velocity, temperature, thermal pressure and energy density,
respectively, $\mathbf{I}$ identity matrix, $\gamma$ is the heat  
capacity ratio, $L_{\mathrm{rad}}(Z,T)$
is the energy loss rate, and $\mathbf{q}$ is the heat flux due to
electron conduction (see sub-section 2.2). The system is closed with
an ideal gas law given by $E=\rho \vert \mathbf{u}\vert^2/2+ P/(\gamma-1)$.
To find the energy loss rate, we use a tabulated cooling function from
the freely available {\sc chianti} database \citep{Dere1997, Landi2006}. As we show in
Figure~\ref{f:cooling}, we have constructed a cooling function with a
range of metallicities of 0.1--30 $Z_{\odot}$.  

The computational domain is a cube of 140 pc on a side, discretised by
$256^3$ cells in a uniform grid, yielding a resolution of 0.5469
pc\footnote{We have tested that the resolution is sufficient in terms of the numerical convergence of the overall features of the simulations. However some details of the
    flow and the exact values of the luminosities still
    depend on the resolution; see the discussion at the end of Section
    3.}.
From the number of massive stars, one can estimate the mass of the
star cluster ($\sim 3500$~M$_\odot$ using Starburst99,
\citet{Leitherer1995}. In our model we did not consider the total mass of
the paternal cloud, but we selected the size of the simulation box so
that it contains a typical superbubble of radius ($\sim$50 pc, i.e.
DEM\,L\,50 and DEM\,L\,152).

\renewcommand{\thefootnote}{\alph{footnote}}

\begin{table*}
\begin{minipage}{\columnwidth}
\centering
\caption{Parameters of the simulations.}
\label{table:intro1}
\begin{tabular}{l*{4}{c}}
\hline
Model & SN locus & SN detonation & Metallicity & Thermal \\
 & & time & & Conduction \\
              & $[\mathrm{pc}]$ & $[10^{5}~\mathrm{yr}]$ &
              $Z_\odot$ &  \\
\hline \hline
sr0tsn2z0.3 & 0  & 2   & 0.3      & no \\
sr0tsn3z0.3 & 0  & 3   & 0.3      & no \\
sr0tsn5z0.3 & 0  & 5   & 0.3      & no \\
sr0tsn2zv    & 0  & 2   & variable\footnotemark[1] & no \\
sr0tsn3zv   & 0  & 3   & variable\footnotemark[1] & no \\
sr0tsn5zv   & 0  & 5   & variable\footnotemark[1] & no \\
sr0tsn7.5zv  & 0  & 7.5 & variable\footnotemark[1] & no \\
sr0tsn2zvC   & 0  & 2   & variable\footnotemark[1] & yes \\
sr0tsn5zvC   & 0  & 5   & variable\footnotemark[1] & yes \\
sr5tsn5zv    & 5  & 5   & variable\footnotemark[1] & no \\
sr10tsn5zv   & 10 & 5   & variable\footnotemark[1] & no \\
\hline
\end{tabular} 
\footnotetext[1]{For these models the metallicity of the ISM is
   $0.3~Z\odot$, $3~Z_\odot$ for the stellar winds, and $10~Z_\odot$
   for the SN ejecta.}
\end{minipage}
\end{table*}

The simulations include 15 stellar wind sources placed randomly within
the cluster radius ($R_c=10~\mathrm{pc}$, and the distribution is
the same for all the simulations). 
The stellar winds are imposed in spherical regions of radius $R_\mathrm{w}=6.03\times
10^{18}$cm (1.95 pc), corresponding to 5 pixels of the grid, and have a
temperature $T_\mathrm{w}=10^5$~K. All the stars have the same mass-loss
rate, $\dot{M}_\mathrm{w}=10^{-6}$~M$_\odot$ yr$^{-1}$, and wind velocity
of $v_\mathrm{w}=1500$~km~s$^{-1}$. We turned on the stellar winds at
the beginning of the simulation. The rest of the computational domain
was initially filled by a homogeneous enviromnent with temperature
$T_\mathrm{0}=10^4$~K and density $n_\mathrm{0}=2$ cm$^{-3}$.  

We impose a SN inside the bubble formed by the winds at four different
times:  $2,~3,~5 $ and $7.5\times 10^5~\mathrm{yr}$ (each
corresponding to a different model). 
These times were chosen to control the distance from the site of
  the supernova to the supperbuble shell.
  Since we do not follow the evolution of the star cluster that
  produces the shell \citep[as][]{Rogers2014}, these times are related
  to the superbubble dynamical age that would be observed and not with
  the star cluster age. The superbubble dynamical age is smaller than
  that of the stars because it does not include the time needed to
  clear up the material between the stars and form a common bubble.
While in most of the models the supernova is placed at the centre of
the star cluster, we have included two 
off-centre models: one in which the SN is placed $5~\mathrm{pc}$
from the centre, and one where it is near the edge of the bubble
($10~\mathrm{pc}$ from the centre). The supernova explosion was
  imposed by injecting a total energy of 1$\times10^{51}$~erg and
  2~M$_\odot$ of mass in a region with a radius of 2~pc. Half of this
  energy was injected as kinetic energy (with velocity following an
  increasing linear profile with radius, and 
  constant density and temperature inside the imposition region),
  and the rest is thermal energy \citep{Toledo2014}. 

To explore the effect of metallicity  we have performed 
some runs with a homogeneous metallicity for all the three components
(the ISM, the stellar winds and the SN) and some models with a
different metallicity for each of these components.
In the homogeneous metallicity models we have used a metallicity of
$0.3~Z_{\odot}$. For the variable metallicity models, following \citealt{Silich2001}, we
use $Z_\mathrm{ISM}=0.3~Z_{\odot}$ for the ISM,
$Z_\mathrm{wind}=3.0~Z_{\odot}$ for the mass injected in the form of
winds, and $Z_\mathrm{SN}=10~Z_{\odot}$ for the SN ejecta.
We have also included thermal conduction in two of the models.

The parameters of the simulations are listed in Table \ref{table:intro1}.
As can be seen from the table, we named the models to reflect the
parameters used: the number after `sr' corresponds to the
location of the SN in pc; it is followed by `tsn' and another 
number to indicate the time of the SN detonation since the winds
sources were turned on (in units of $10^5~\mathrm{yr}$); next
there is the letter `z' followed by the number $0.3$ for the uniform
metallicity models or the letter `v' for the variable metallicity
runs; for the models with thermal conduction a letter `C' is appended
at the end of the model name. 


\subsection{Adding the effect of metallicity to the cooling}

The cooling in the code is added as a source term after updating the
hydrodynamic variables. At the end of each timestep, we estimate the cooling by
interpolating a tabulated cooling curve which is, for a given metallicity,
a function of the temperature. The energy loss is then subtracted to the
internal energy of each cell at every timestep.

For the runs with a uniform metallicity this is a simple linear
interpolation (in temperature) of a single table that is generated by
the {\sc chianti} database. For the runs with varying
metallicity we created a series of tables for metallicities in the range of
$0.1$--$30~Z_\odot$; these are plotted in  Figure~\ref{f:cooling}. 
Along with the gas-dynamic equations
(eqs. \ref{eq:conmass}--\ref{eq:conene}) we consider the metallicity
$Z$ as a passive scalar by including an extra equation of the form 
\begin{equation}
\label{eq:conmet}
\frac{\partial Z \rho}{\partial t}+\mathbf{\nabla}\cdot(Z \rho \mathbf{u})=0.
\end{equation}
Using the metallicity value at each cell we do a bi-linear
interpolation (with metallicity spaced linearly and temperature
  logarithmically) to estimate the cooling to be applied there. 
For the ISM the metallicity is set to $0.3~Z_\odot$ at the start of the simulation.
For the winds and supernovae the gas is injected into the simulation either with
$Z_{\mathrm{wind}}=Z_\mathrm{SN}=0.3~Z_\odot$ (for the homogeneous models) or 
$Z_{\mathrm{wind}}=3~Z_\odot$ and $Z_\mathrm{SN}=10~Z_\odot$ for the inhomogeneous models.

The average metallicity in each region can be calculated (using
\citealt{Silich2001}) as:

\begin{equation}
\label{eq:metallicity}
\tilde{Z}=\frac{M_\mathrm{z,ej}+M_\mathrm{z,ism}}{M_\mathrm{ej}+M_\mathrm{ism}}
\end{equation}
where, $M_\mathrm{z,ej}$ and $M_\mathrm{z,ism}$ are the masses of the
metallic ejecta (by winds and/or SN) and swept up interstellar gas, respectively, while
$M_\mathrm{ej}$ and $M_\mathrm{ism}$ are the total masses of the
ejected and swept up interstellar gas.

In general, the most metal enriched regions are found behind
the contact discontinuity that separates the main and reverse
shocks. Even though some mixing occurs at the interface (mainly due
to hydrodynamical instabilities and/or turbulence), since the swept up
ISM mass is larger than the ejected mass the metallicity of the shell
remains close to that of the ISM.

\subsection{Thermal conduction}
In order to include the effect of thermal conduction by free
electrons in our numerical simulation, we add a heat flux term 
($\mathbf{\nabla} \cdot \mathbf{q}$) in the right hand side of the
energy equation (\ref{eq:conene}). 

The heat conduction due to collisions with free electrons in a plasma
is given by the classical \citet{Spitzer1962} law:
\begin{equation}
\label{eq:heatflux}
\mathbf{q}=-k\mathbf\nabla T,
\end{equation}
where $k$ is the thermal conductivity given by
\begin{equation} 
\label{eq:thermcond}
k=\beta~T^{5/2}.
\end{equation}
where, for a fully ionized hydrogen plasma $\beta\approx 6\times
10^{-7}~\mathrm{erg~s^{-1}~K~cm^{-1}}$ \citep[see][]{Spitzer1962}. The
result relies on the assumption that the mean free path is small
compared to the scale-length of temperature variations  ($\lambda \ll T/|\nabla T|$).

When the mean free path of the electrons is comparable or larger than
temperature scale-length the heat flux saturates. In this regime
the heat flux can be estimated by the local sound speed
($c_\mathrm{s}$) and pressure ($P$), as described by  \citet{Cowie1977}:
\begin{equation}
\label{eq:heatfluxsat}
q_\mathrm{sat}=5 \phi_\mathrm{s} c_\mathrm{s} P,
\end{equation}
where $\phi_\mathrm{s}$ is a factor of order unity (we have used
$\phi_\mathrm{s}=1.1$).

At every timestep we compute the heat fluxes in the classical and the
saturated regimes, keep the smaller one and introduce its
divergence as a source term to the energy equation. We have to mention
that the thermal conduction timescale is smaller than the hydrodynamic
one determined by the standard CFL condition. For this reason we
apply a sub-stepping method to include the source term (we take on the
order of 100 sub-steps to integrate the source term for each hydrodynamical step).

\subsection{X-ray emission coefficients}

\begin{figure}
\centering
\includegraphics[width=\columnwidth]{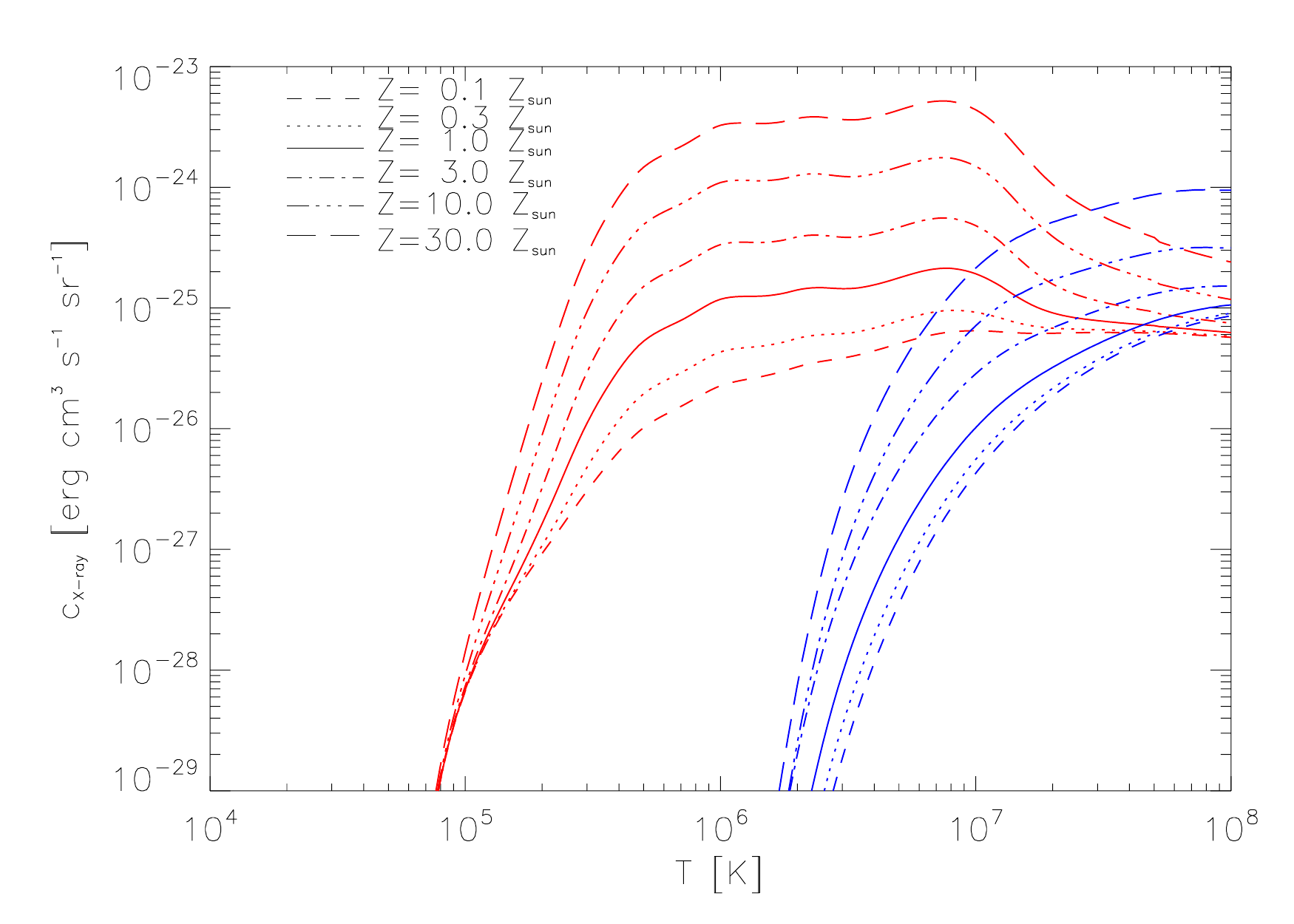}
\caption{Thermal soft (0.1--2 keV, in red lines) and hard (2--10 keV, in blue 
lines) X-ray emission coefficients for a range of metallicities 
between 0.1 to 30~Z$_{\odot}$.}
\label{f:coeftot}
\end{figure}

We take the output from the hydrodynamical simulations to estimate
the X-ray luminosity in all the models. We consider that the emission
coefficient in the low-density regime is
$j_{\nu}(n,Z,T)=n_\mathrm{e}^2\chi(Z,T)$, where $n_\mathrm{e}$ is the
electron density, and $\chi(Z,T)$ is a function of the the temperature
($T$) and the metallicity ($Z$). For a given metallicity, the function
$\chi$ can be computed and integrated over an energy band using the
{\sc chianti} atomic database and its associated IDL software
\citep{Dere1997,Landi2006}. 
We have computed $\chi(Z,T)$ for various metallicities
($Z=0.1$, $0.3$, $1$, $3$, and $10~Z_\odot$), using the ionisation
equilibrium model by \citet{Mazzota1998}, over a range of temperatures
from $10^4$ to $10^9~\mathrm{K}$. The emission coefficients were
integrated over two energy bands: soft X-rays
($0.1$--$2~\mathrm{keV}$), and hard X-rays ($2$--$10~\mathrm{keV}$).
The result is a two dimensional table of coefficients that is function
of temperature and metallicity. 
Figure~\ref{f:coeftot}  show the thermal 
soft (red lines) and hard (blue lines) X-ray emission coefficients, respectively, 
as function of temperature for several metallicities. 

From the results of the simulations we obtain the density, temperature and
metallicity in every computational cell and perform a bilinear
interpolation to get $\chi$, and then use it to determine the emissivity for the two
energy bands. The contribution of all cells are then added
to compute the total X-ray luminosity both for soft and hard X-rays.


\section{Results}

\begin{figure*}
\centering
\includegraphics[width=\textwidth]{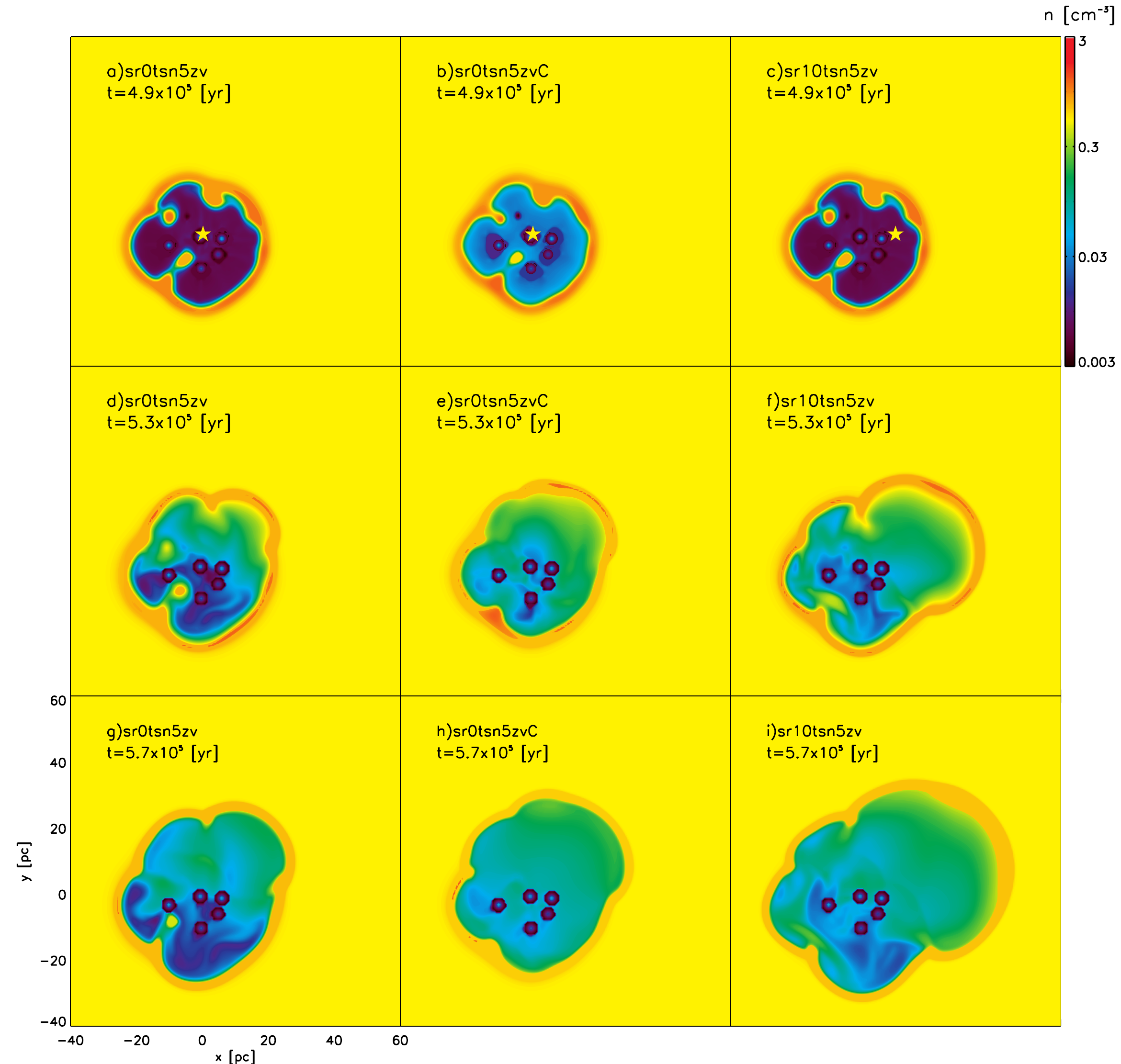}
\caption{
Density maps  at different
stages of the evolution and for different models. The columns
correspond to three distinct models: in the panels of the left column,
the SN explosion occurs at the centre of the cluster (model sr0tsn5zv;
panels a, d and g); those of the central column show the same model
but including thermal conduction (sr0tsn5zvC; panels b, e and h); and
in those of the right column the SN explosion occurs 10 pc off-centre
(sr10tsn5zv; panels c, f and i). For all models shown here, the SN
explosion occurs at $t=5\times10^5$ yr and the metallicity of the gas
varies across components, as discussed in the text. The rows
correspond to three relevant evolutionary stages: just before the SN
explosion (top row), at the peak of X-ray luminosity (middle row), and
once the luminosity has approximately returned to its original level
(bottom row). The position of the SN is marked with a star in the
panels of the top row. The spatial scale is the same across all panels
and is shown in panel g.
}
\label{f:Den}
\end{figure*}

\begin{figure*}
\centering
\includegraphics[width=\textwidth]{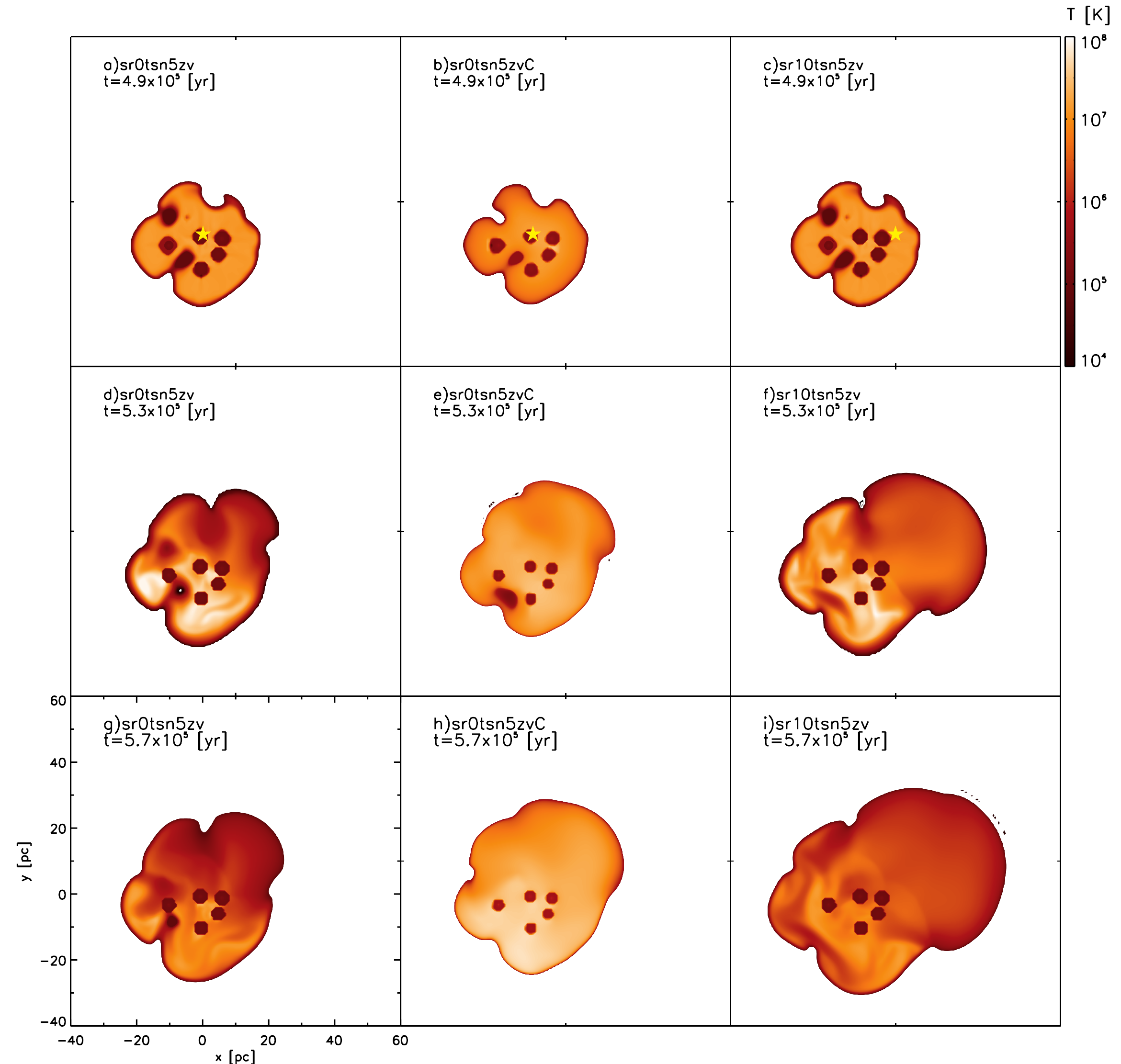}
\caption{
Temperature maps, corresponding to the same panels as in Figure~\ref{f:Den}.
}
\label{f:Temp}
\end{figure*}

The colour maps of Figures \ref{f:Den} and \ref{f:Temp} show the
  density and temperature at three different evolutionary times. These were
  chosen to show the effect of the SN explosion in the X-ray
  emission (see the next section). We present a time slightly
  before the SN explosion (top row), at the peak  
of luminosity after the explosion (middle row) and once the total
luminosity has diminished back to a value near its pre-SN level
(bottom row). The columns correspond to three different models: in the
left and central columns the SN ocurrs at the centre of the cluster,
without thermal conduction, and with thermal conduction respectively,
and in the rightmost column the SN is 10 pc off-centre (the position
of the SN is indicated by a star in the top row). Following the time
sequence in the columns of this figure, it can be seen that the SN
ejecta reach the edge of the wind bubble and push it further into
the ambient medium. Due to the particular position of the stars in
these models, the gas distribution inside the wind bubble favors the
expansion of the SN eject towards the upper right corner of the
simulation box, and thus the blowout is more pronounced in this
direction, the effect being larger if the SN explodes off-centre (at
the edge of the star cluster; see the rightmost panels).

\subsection{Soft X-ray emission}

Following the pressure driven model discussed by \citet{Chu1990}, the
soft X-ray luminosity can be estimated from  
\begin{equation}
\label{eq:chu}
L_\mathrm{X}=3.29 I(\tau) \xi L^{33/35}_{37} n^{17/35}_0 t^{19/35}_6 \,\, \rm{[erg~s^{-1}]},
\end{equation}
where
\begin{equation}
\label{eq:Itau}
I(\tau)=\frac{125}{33}-5\tau^{1/2}+\frac{5}{3}\tau^3-\frac{5}{11}\tau^{11/3},
\end{equation}
with
\begin{equation}
\label{eq:tau}
\tau=0.16L^{-8/35}_{37}n^{-2/35}_0 t^{6/35}_6,
\end{equation}
$\xi$ is the gas metallicity, $L_{37}=L_\mathrm{w}/10^{37}$ where
$L_\mathrm{w}$ is the mechanical luminosity of the cluster (in
$\mathrm{erg~s^{-1}}$), $n_0$ the interstellar medium density
 and $t_6$ is the cluster lifetime in Myr.
In all the models presented here, the mechanical energy injected by
the winds is $1.1 \times10^{37}~\mathrm{erg~s^{-1}}$.
With this mechanical energy the total X-ray luminosity for the
stellar wind contribution is $\sim 10^{33}~\mathrm{erg~s^{-1}}$ (see
eq.~\ref{eq:chu}, and also \citealt{Chu1990}), for an interstellar medium
density of 2 cm$^{-3}$ and metallicity of 0.3~$Z_{\odot}$ after an evolution time of
$2\times10^5$~yrs. 

\begin{figure}
\centering
\includegraphics[width=\columnwidth]{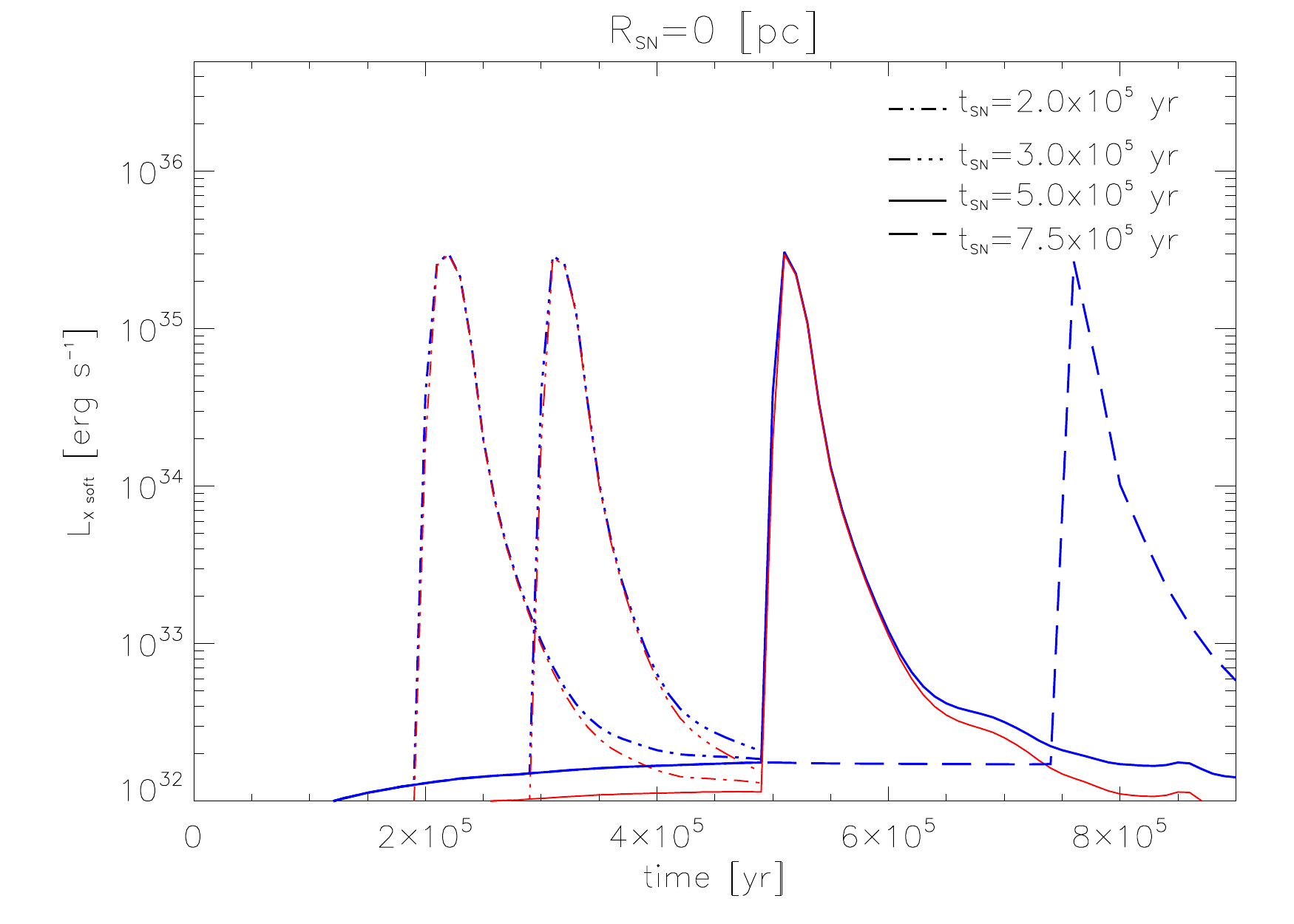}
\caption{Evolution of the total soft X-ray luminosities for models
  with the supernova explosion occuring at the centre of the star
  distribution. The red lines are the models with homogeneous
  metallicity (for a supernova event at t=2, 3, 5, and
  7.5$\times$10$^5$~yr dash-dotted, solid, dot-dash-dotted and dashed 
lines respectively). The blue lines are the models with different
metallicities ($Z_{\mathrm{ISM}}=0.3$~$Z_\odot$, $Z_{\mathrm{wind}}=3$~$Z_\odot$ and
$Z_{\mathrm{SN}}=10$~$Z_\odot$ for the interestellar medium, stellar wind and
supernova explosion, respectively,  for SN explosions at t=2, 3 and
5$\times$10$^5$~yr).}
\label{f:lumXsoftR0}
\end{figure}

We have computed the soft X-ray emission for all the models at $10^{4}$ yr intervals.
Figure~\ref{f:lumXsoftR0} shows the evolution of the total soft X-ray
luminosities for all models without thermal conduction and with the SN placed
at the centre of the star cluster. The red lines show the models with
uniform metallicity, and the blue ones with variable metallicity.

The X-ray luminosity before the supernova event is in agreement with
the  value predicted by \citet{Chu1990}. Shortly after the supernova
explosion the luminosity increases dramatically. We calculated the
time interval in which the soft X-ray luminosity remains above
10$^{34}$, 10$^{35}$  and 10$^{36}$~erg s$^{-1}$ ($\Delta t_{\mathrm{s},34}$, $\Delta
t_{\mathrm{s},35}$ and  $\Delta t_{\mathrm{s},36}$, respectively). The
maximum luminosity achieved and these time intervals are shown in Table~\ref{table:soft}.
\begin{table}
\caption{Maximum soft X-ray luminosity and time intervals in which the
  soft X-ray emission remains above 10$^{34}$, 10$^{35}$ and 10$^{36}~\mathrm{erg~s^{-1}}$.}
\label{table:soft}
\begin{center}
\begin{tabular}{l*{4}{c}}
\hline
      Model   & L$_{\mathrm{max,soft}}$&  $\Delta t_{\mathrm{s},34}$      &
      $\Delta t_{\mathrm{s},35}$    &  $\Delta t_{\mathrm{s},36}$   \\
 & [erg s$^{-1}$]& [$10^4$yr] &  [$10^4$yr] & [$10^4$yr]  \\ \hline \hline
sr0tsn2z0.3  & 2.98$\times10^{35}$ & 6.42 &       3.55  &    ---\\
sr0tsn3z0.3  & 2.83$\times10^{35}$ & 5.60 &       3.05  &    ---\\
sr0tsn5z0.3  & 2.98$\times10^{35}$ & 6.01 &       2.84  &    ---\\ \hline
sr0tsn2zv    & 3.01$\times10^{35}$ & 6.72 &      3.63  &    --- \\
sr0tsn3zv    & 2.92$\times10^{35}$ & 5.96 &      3.13  &    --- \\
sr0tsn5zv    & 3.08$\times10^{35}$ & 6.34 &      2.92  &    --- \\
sr0tsn7.5zv  & 2.68$\times10^{35}$ & 6.13 &      2.90  &    --- \\ \hline
sr0tsn2zvC   & 3.65$\times10^{35}$ & 8.20 &      4.30  &    --- \\
sr0tsn5zvC   & 4.03$\times10^{35}$ & 9.00 &      3.96  &    --- \\\hline
sr5tsn5zv    & 4.16$\times10^{35}$ & 7.61 &      3.79  &    --- \\
sr10tsn5zv   & 1.23$\times10^{36}$ & 13.81 &     7.27  &    3.02 \\ \hline
\end{tabular}
\end{center}
\end{table}

\begin{figure}
\centering
\includegraphics[width=\columnwidth]{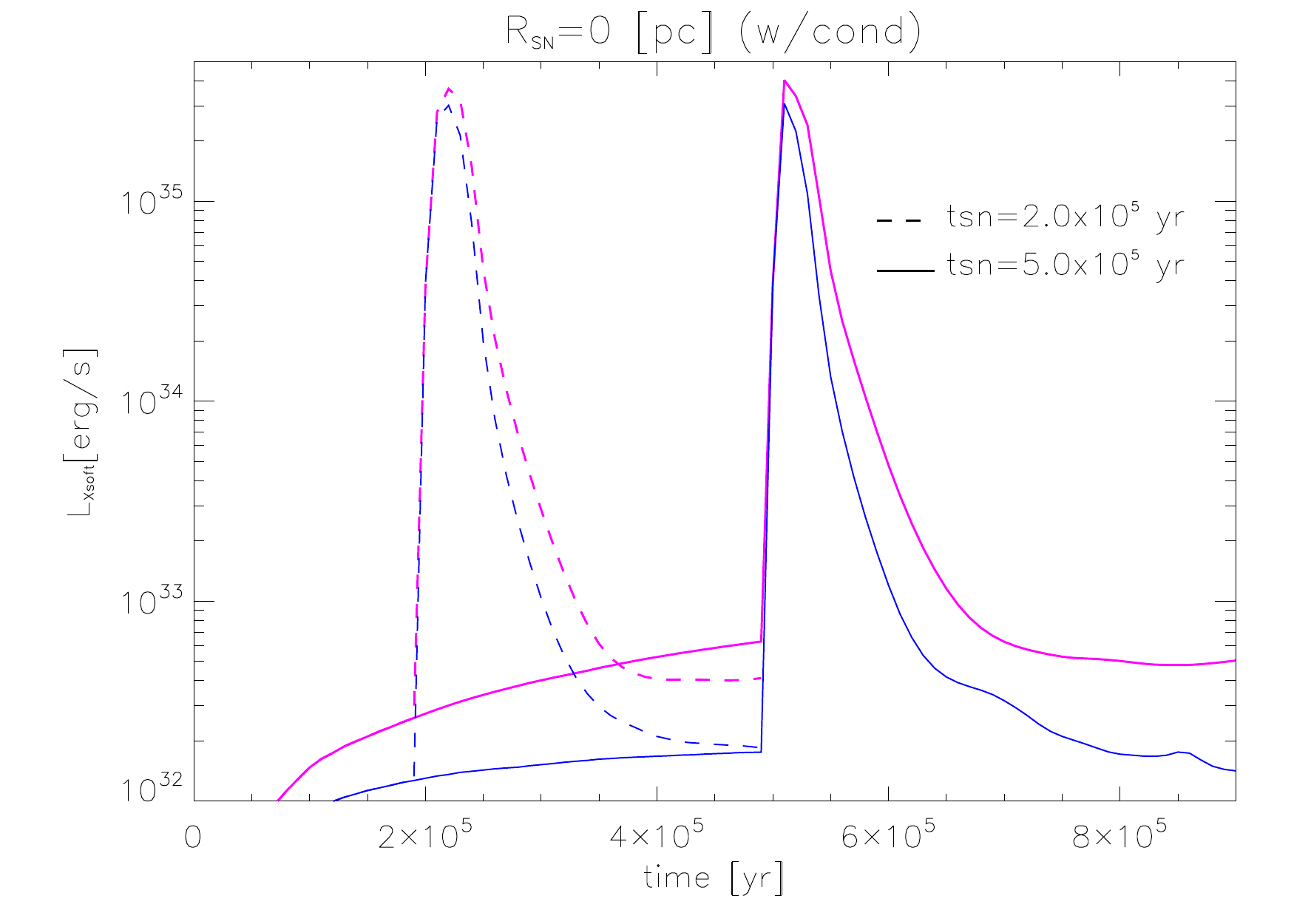}
\caption{Evolution of the total soft X-rays luminosities for models with supernova explosion 
in the centre of the star distribution with inhomogeneous metallicity. The magenta line are the models with 
thermal conduction and the blue lines are the models without thermal conduction process 
(for supernova event at t=2, and 5 $\times$10$^5$~yr dashed and solid lines, respectively).}
\label{f:lumXsoftR0C}
\end{figure}

From Figure~\ref{f:lumXsoftR0}, we can see that the general shape of
the luminosity curve after the SN explosion is quite similar in
all the models. All models reached the same maximum luminosity of
$\sim 3\times10^{35}$~erg$^{-1}$, and have $\Delta t_{\mathrm{s},34}\sim 6 \times 10^{4}$~yr
and $\Delta t_{\mathrm{s},35}\sim$3$\times10^{4}$~yr.
In these models, in which the SN explosion occurs at the centre of the
stellar distribution, luminosities above 10$^{36}$~erg s$^{-1}$ are
never reached.
Our results show only small differences in the soft X-ray emission between
models with uniform metallicity and those with variable metallicity.
The similarity of the emission across the models indicates that the
emission is dominated by swept up ISM material. This is partly because
the high metallicity gas (winds and SN) is kept at a temperature too
high for thermal soft X-rays to be important. 
Thermal conduction allows energy transport with reduced bulk motions,
resulting in a denser inner region. \citet{Weaver1977}
estimated a significant increase in the soft X-ray luminosity (up to 2
orders of magnitude) with respect to models without thermal conduction.
Figure~\ref{f:lumXsoftR0C} displays the evolution of the soft X-rays
luminosities for the 2 models with thermal conduction (sr0t2e5zvC and
sr0t5e5zvC, magenta lines), and their counterpart without thermal
conduction (sr0t2e5zv and sr0t5e5zv, blue lines).
As can be seen, thermal conduction does increase the maximum luminosity of the
models, but only by a factor of $\sim1.25$, and the luminosity returns to a value a factor of $2$
larger after the SN explosion.
The time of emission above $10^{34}~\mathrm{erg~s^{-1}}$ also
increases by a similar factors of $1.2$ and $1.6$ for the SN imposed after
$2\times 10^5~\mathrm{yr}$ and $5\times 10^5~\mathrm{yr}$,
respectively. Approximately the same time-span increase is found for
emission above $10^{35}~\mathrm{erg~s^{-1}}$; see
Table~\ref{table:soft}.
These increments in the time interval with emission above $10^{34}$ and/or
$10^{35}$~erg s$^{-1}$ will enhance the chance of such luminosities
being observed. 

We can see that the inclusion of different metallicities and/or thermal
conduction induces only small discrepancies in the soft X-ray emission.

From this models it is clear that the supernova explosions are a
crucial ingredient for the thermal X-ray emission.
The presence of a SN can explain the extra
X-ray luminosity observed in several superbubbles. 
However, when the supernova event occurs in the centre of the cluster,
the soft X-ray luminosity only reaches a few times 10$^{35}$~erg
s$^{-1}$, still falling short of some of the observed values 
\citep[e.g., those of][]{Jaskot2011}. We find that models with a centred
explosion seem to still be underluminous.

\begin{figure}
\centering
\includegraphics[width=\columnwidth]{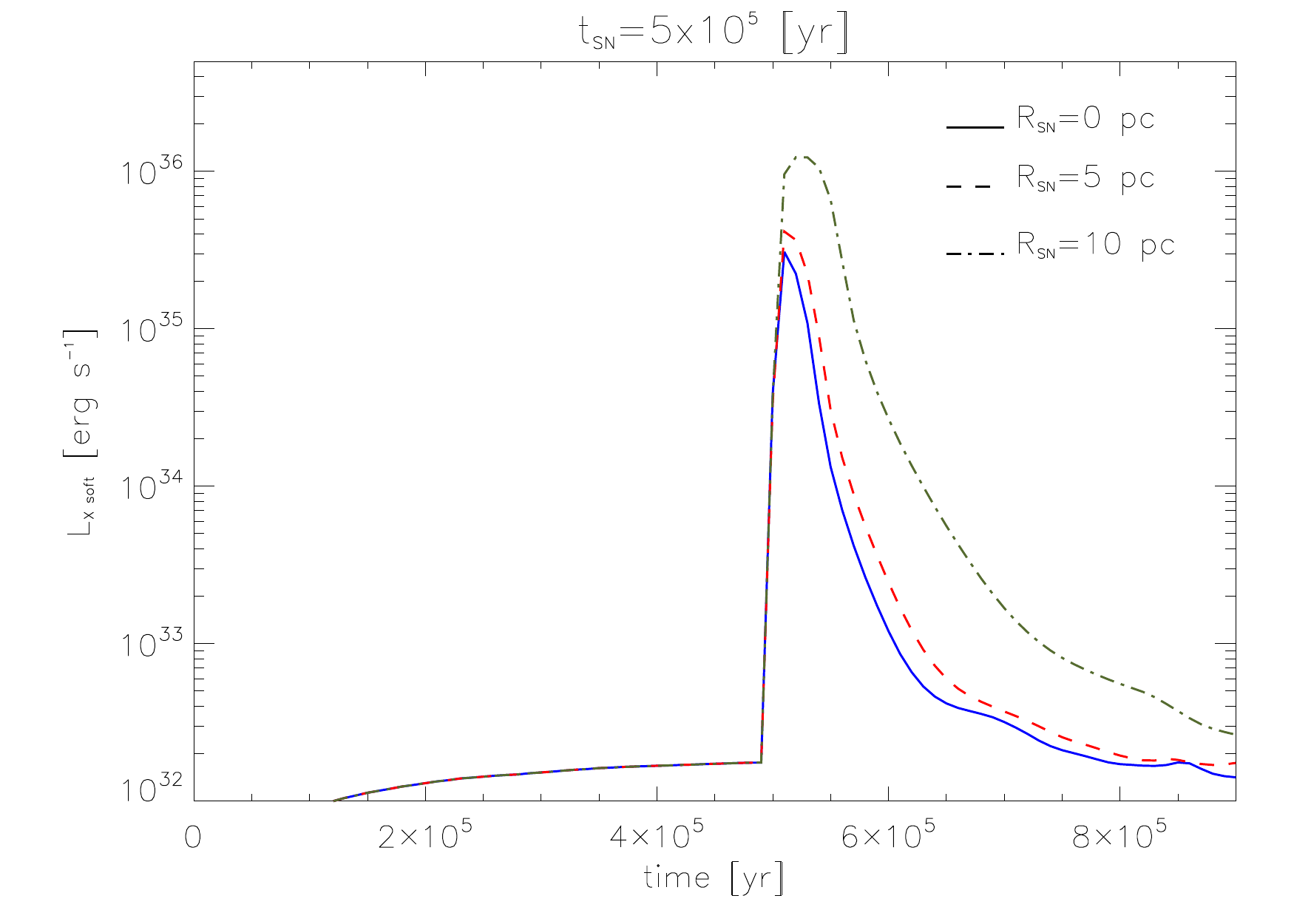}
\caption{Evolution of the total soft X-rays luminosities for models with supernova explosion at 
t=5$\times10^5$~yr, and with inhomogeneous metallicity. The blue,red and olive lines are the model with 
supernova event in R=0, 5 and 10~pc, respectively.}
\label{f:lumXsoftR}
\end{figure}

A possibility that results in luminosities above $10^{36}$~erg~s$^{-1}$ is to
place the SN at a distance from the centre of the star cluster. For these
reason we have included models sr5tsn5zv and sr10tsn5zv where the SN occurs
at $R_\mathrm{SN}=5$ and $10$ pc from the star cluster centre, respectively,
both at $t_\mathrm{SN}=5\times10^5$~yr. Figure~\ref{f:lumXsoftR} shows the
soft X-ray luminosities for these two models compared to a model with a
SN placed at the cluster centre (sr0tsn5zv).


We can see that the X-ray luminosities increase when the SN explodes
closer to the edge of the bubble.
The maximum soft X-ray luminosity increases by a factor of $\sim 1.25$
between the model with supernova explosion at $R_\mathrm{SN}=0$ and
$R_\mathrm{SN}=5$~pc and a factor of $\sim 4$  when the SN explodes
near the cluster radius  ($R_\mathrm{C}=10$~pc), reaching  a maximum
luminosity of $L_\mathrm{max}=1.23\times10^{36}$~erg~s$^{-1}$.

For model sr10tsn5zv, the only one  that reached
$10^{36}~\mathrm{erg~s^{-1}}$, the time interval spent 
above $10^{36}$~erg~s$^{-1}$ was $30$ kyr. In addition,
this last model  predicts a time spent above $10^{35}$~erg~s$^{-1}$ 
of $72$~kyr, and one   above $10^{34}~\mathrm{erg~s^{-2}}$ of $14$~kyr.
These numbers are $\sim 3$ times larger than those of the model with 
the supernova explosion occurring at the centre of the star cluster. 
  
From these results we can see that, on the one hand, for the case of
the SN at the cluster centre the soft X-ray luminosity increase is not very sensitive to the time at which it is detonated. The luminosity increase is only slightly larger for SN explosions that occur later in the evolution of the superbubble. On the other hand, when the SN is off-centre the luminosity increase depends considerably on the distance to the centre of the cluster.

\begin{figure*}
\centering
\includegraphics[width=\textwidth]{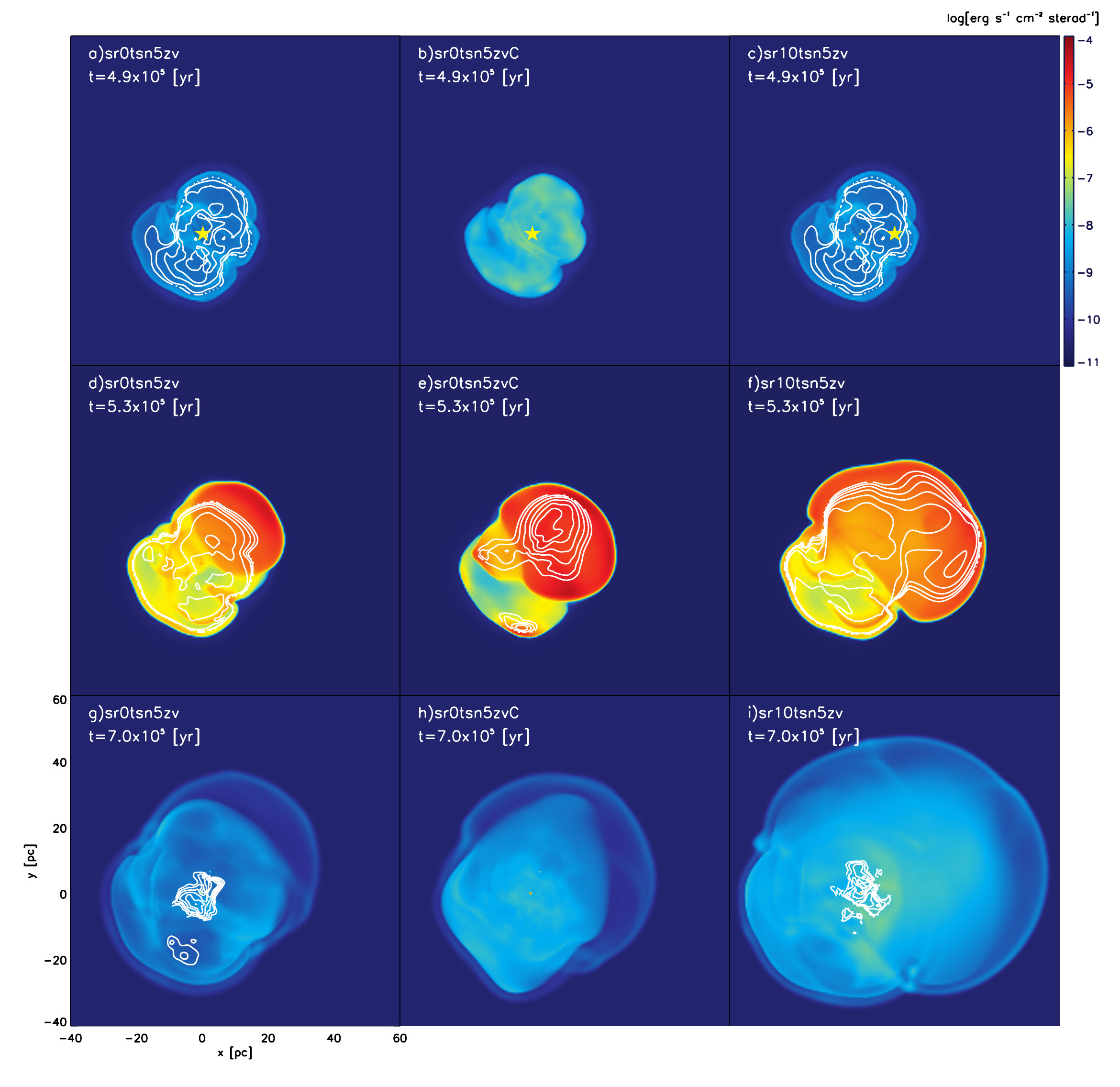}
\caption{
Comparison of the X-ray emissivity in the soft (color maps, with the
shown logarithmic scale given in erg s$^{-1}$ cm$^{-2}$ sterad$^{-1}$)
and hard (contours, logarithmically-spaced levels from $10^{-11}$ to
$10^{-8}$ erg s$^{-1}$ cm$^{-2}$ sterad$^{-1}$) bands at the three different
stages shown in Figure \ref{f:Den}. 
}
\label{f:Xrays}
\end{figure*}

The color maps of Figure \ref{f:Xrays} show the soft X-ray
  emissivity in the same layout as in Figure \ref{f:Den}. Note that
  the blowout region (located at the upper right corner of the
  simulation box) provides the largest contribution to the luminosity
  increase seen 30 kyr after the explosion (middle panels). By 200~kyr after the
explosion (bottom panels), the emissivity has almost returned to
values comparable to its pre-SN level; however, due to the now larger
X-ray emitting volume, the luminosity remains slightly above the
original level (see Figure \ref{f:lumXsoftR}).

An interesting exercise is to compare the predicted luminosity statistics
of our simulations with those of observed superbubbles. For this purpose
we have taken a sample of 26 bubbles with luminosities greater than
$10^{34}$~erg~s$^{-1}$  from the literature \citep{Oey1996, Jaskot2011,
Reyes2014, Dunne2001}. 
Out of these, 18 have luminosities above $10^{35}$~erg~s$^{-1}$  while only 3
are observed with L$_\mathrm{soft} > 10^{36}$~erg~s$^{-1}$ . We can use our
numerical results to predict how many bubbles out of these 26 should have
luminosities above these two levels. In order to do this, we computed, from
Table 2, the ratios of the times spent above these levels to the overall time
where luminosity is above $10^{34}$~erg~s$^{-1}$  for model sr10tsn5zv
(the only one that reaches $10^{36}$~erg~s$^{-1}$). We find that 
$\Delta t_{\mathrm{s},36}/ \Delta t_{\mathrm{s},34}\sim 52.6$ \% and 
$\Delta t_{\mathrm{s},35}/\Delta t_{\mathrm{s},34}\sim 21.9$ \%.
Thus, assuming that all bubbles in this observed  sample reach a luminosity of
10$^{36}$~erg~s$^{-1}$  at some point in their evolution, this model predicts that
about 14 should have a luminosity above 10$^{35}$~erg~s$^{-1}$  while about
6 should be observed above 10$^{36}$~erg~s$^{-1}$. Though the values do not
coincide exactly, they reasonably match the luminosity ratios of 
L$_{36}$/L$_{34}\sim$ 12 \% and  L$_{36}$/L$_{35}\sim$ 70 \% observed in 
the sample. Here, L$_{n}$ is a soft X-ray luminosity that is greater than or equal to 
$10^{n}$~erg~s$^{-1}$.  

There could be several explanations for this difference. For 
one, it is hard to judge whether this small sample of superbubbles is 
representative of the general population, and thus some variability can 
be expected in the statistics. At the same time, our numerical results 
suggest that the position of a supernova explosion occurring inside the 
bubbles determines whether a luminosity of 10$^{36}$~erg~s$^{-1}$  is reached at all 
during their lifetimes. Thus, if not all of the observed bubbles have 
had off-centre SN explosion, it is to be expected that fewer of them 
would be observed above 10$^{36}$~erg~s$^{-1}$  than what our models predict.

\subsection{Hard X-ray emission}

Hard X-ray emission is produced in the hottest regions inside the
bubble where individual winds interact, when the gas flow is faster than a 
$\sim 1000$ km~s$^{-1}$, as well as during the early
stages of the SN remnant evolution and where the cluster wind and/or the SN remnant
are heated by the reverse shock. Thus one should expect that
metallicity should have a significant effect on the hard X-ray
emission.

In Table~\ref{table:hard} we show the maximum thermal hard X-ray
luminosity, and the time intervals for which the luminosity remains
above $10^{33}$~erg~s$^{-1}$ ($\Delta t_{\mathrm{h},33}$) and
$10^{34}$~erg~s$^{-1}$ ($\Delta t_{\mathrm{h},33}$).

\begin{figure}
\centering
\includegraphics[width=\columnwidth]{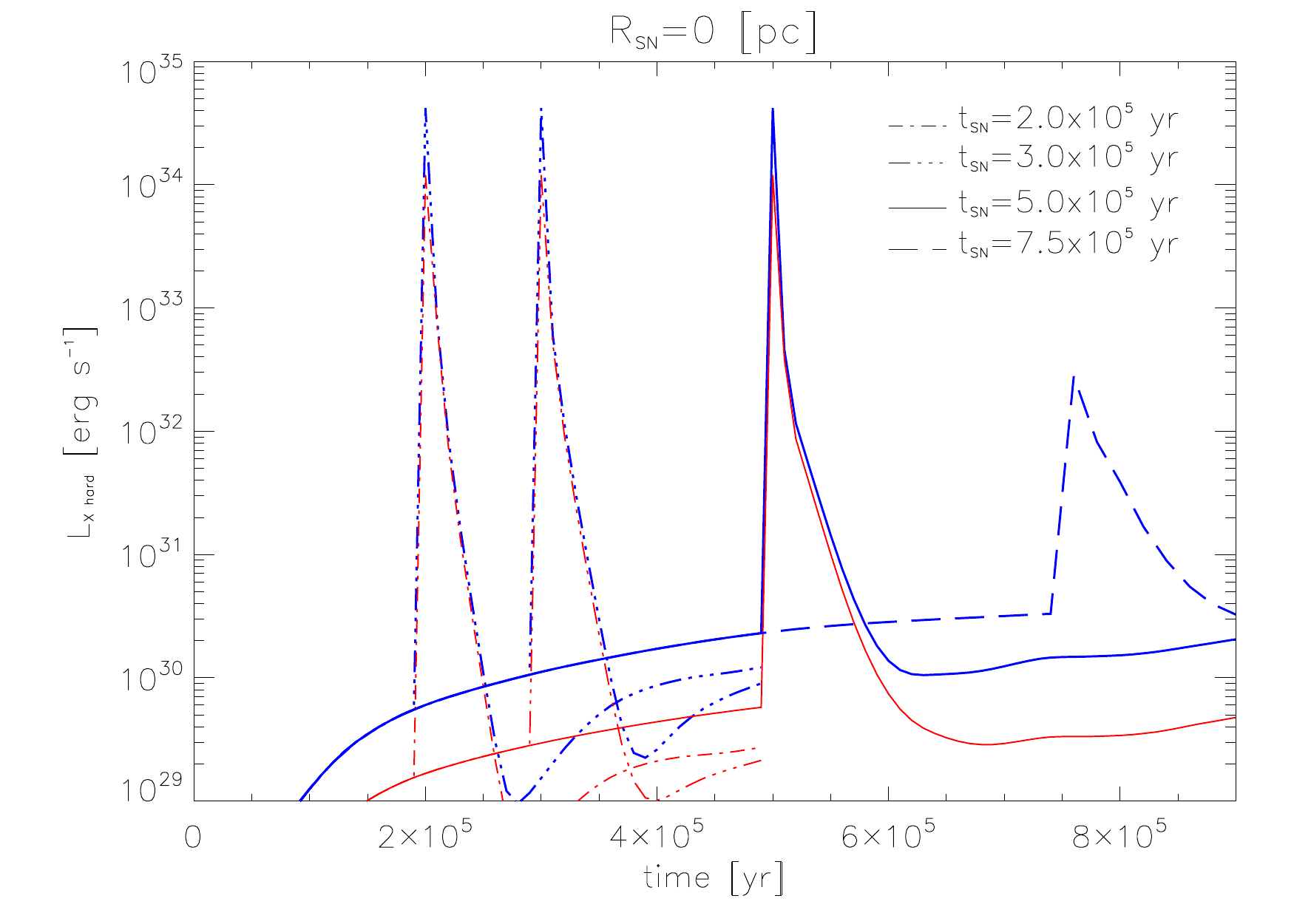}
\caption{Same as Figure~\ref{f:lumXsoftR0} for hard X-rays luminosities.}
\label{f:lumXhardR0}
\end{figure}

Figure~\ref{f:lumXhardR0} shows the evolution of the thermal hard
X-rays luminosity for all models with the SN placed at the centre.
In the models with different metallicities all three components exhibit 
maximum luminosities ($\sim3\times10^{34}$~erg~s$^{-1}$) that are
$\sim 3$ times larger than those of the models with homogeneous metallicity
($\sim 1.0\times10^{34}$~erg~s$^{-1}$). 
We have to mention that the thermal hard X-ray emission produced
inside the star cluster (regions of wind collisions) is
underestimated by the models with homogeneous metallicity due to the
rather low metallicity of the wind sources ($0.3~Z_\odot$).
In the models with the variable metallicity the wind is injected with
a more appropriate metal content ($3~Z_\odot$), thus the hard X-ray
luminosity in these models should be closer to reality.
\begin{table}
\caption{Maximum hard X-ray luminosity and time intervals in which the hard X-ray emission remains above 10$^{33}$ and 10$^{34}~\mathrm{erg~s^{-1}}$.}
\label{table:hard}
\begin{center}
\begin{tabular}{*{4}{c}} \hline
      Model   & L$_{\mathrm{max,hard}}$&  $\Delta t_{\mathrm{h},33}$      &
      $\Delta t_{\mathrm{h},34}$    \\
 & [erg s$^{-1}$]& [$10^4$yr] &  [$10^4$yr]   \\ \hline \hline
sr0tsn2z0.3  & 2.88$\times10^{34}$ & 2.00 &       1.35  \\
sr0tsn3z0.3  & 2.88$\times10^{34}$ & 1.99 &       1.35 \\
sr0tsn5z0.3  & 2.88$\times10^{34}$ & 1.96 &       1.33 \\ \hline 
sr0tsn2zv    & 1.02$\times10^{35}$ & 2.03 &      1.84     \\
sr0tsn3zv    & 1.02$\times10^{35}$ & 2.02 &      1.85      \\
sr0tsn5zv    & 1.02$\times10^{35}$ & 2.00 &      1.83      \\
sr0tsn7.5zv  & 2.81$\times10^{32}$ & 0.00 &      0.00      \\ \hline
sr0tsn2zvC   & 9.90$\times10^{34}$ & 2.03 &      1.84      \\
sr0tsn5zvC   & 9.93$\times10^{34}$ & 2.00 &      1.82      \\\hline 
sr5tsn5zv    & 9.07$\times10^{34}$ & 2.00 &      1.81      \\
sr10tsn5zv   & 9.84$\times10^{34}$ & 3.57 &      1.99      \\ \hline
\end{tabular}
\end{center}
\end{table}


We can see from Table~\ref{table:hard} that the maximum hard X-ray
luminosity is significantly larger in the models with variable
metallicity, typically an increase of $\sim 3.5$ times. The time interval that the emission remains above $10^{33}~\mathrm{erg~s^{-1}}$ is similar, although slightly larger in the models with more
realistic metallicity. In contrast, the time that the emission remains above
$10^{34}~\mathrm{erg~s^{-1}}$ is much larger (a factor of $\sim 4$) than that obtained in the models with homogeneous metallicity.

An important fact to notice is that the ratio of the maximum
luminosities in soft-Xrays to hard X-rays is on the order of $10$.
Vel\'azquez et al. (2013) explored the ratios between soft and hard
X-ray emission in the young star cluster M\,17. This particular
cluster is partially immersed in the cluster parental cloud, and their
models resulted in a ratio of soft to hard X-rays of two orders of
magnitude.
From our models we notice that the time intervals of high luminosity for soft X-rays
($\Delta t_{\mathrm{s},34}$ and $\Delta t_{\mathrm{s},35}$) are
larger than those obtain for hard X-rays
($\Delta t_{\mathrm{h},33}$ and $\Delta t_{\mathrm{h},34}$)

This shows that very young star clusters with a SN event
can produce hard X-ray luminosities that are only an order of
magnitude fainter than the soft X-rays. However, this happens only for
a short interval at the earlier stages of the SN remnant. After that
the hard X-ray emission drops abruptly.

\begin{figure}
\centering
\includegraphics[width=\columnwidth]{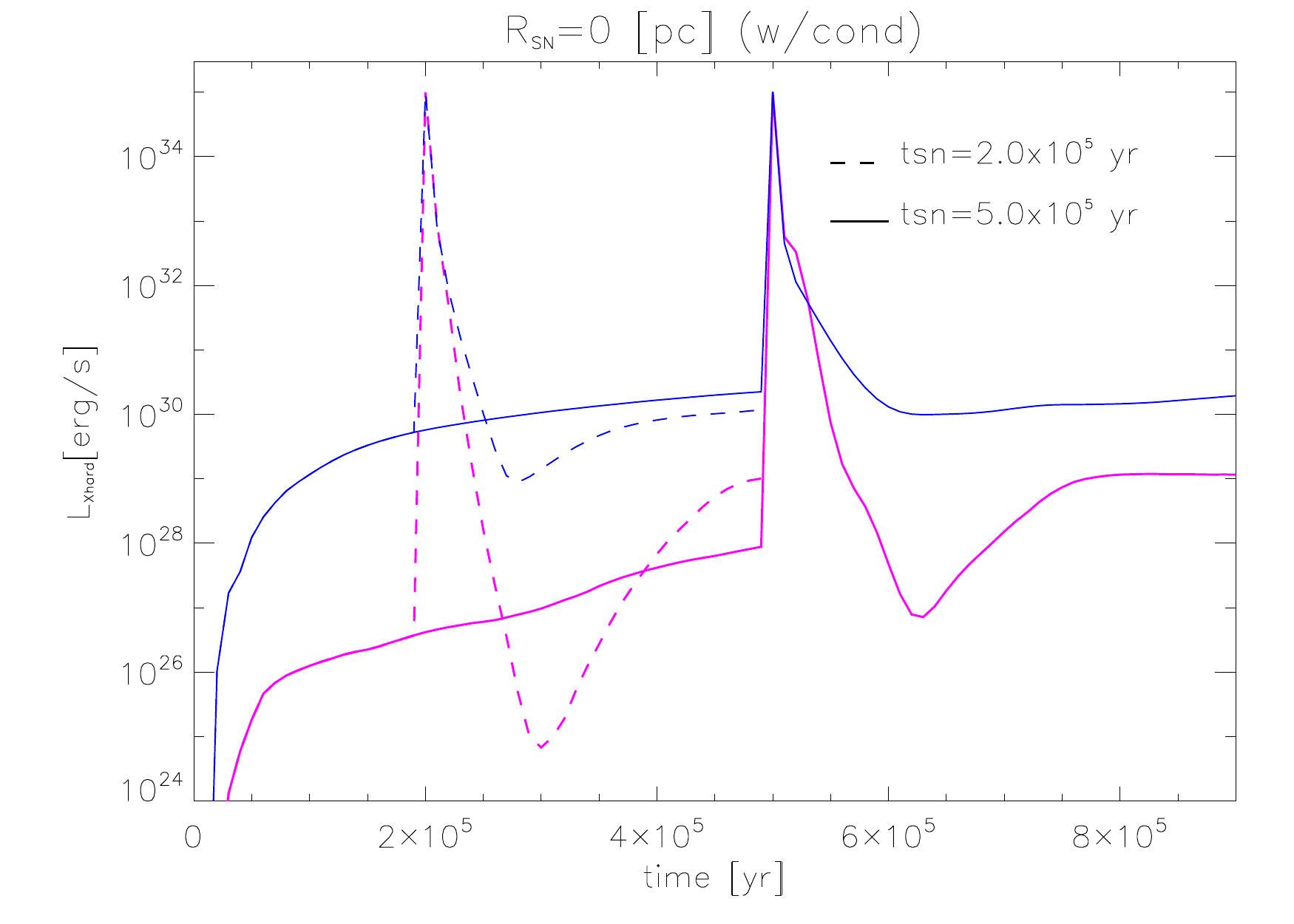}
\caption{Same as Figure~\ref{f:lumXsoftR0C} for hard X-rays luminosities.}
\label{f:lumXhardR0C}
\end{figure}

\begin{figure}
\centering
\includegraphics[width=\columnwidth]{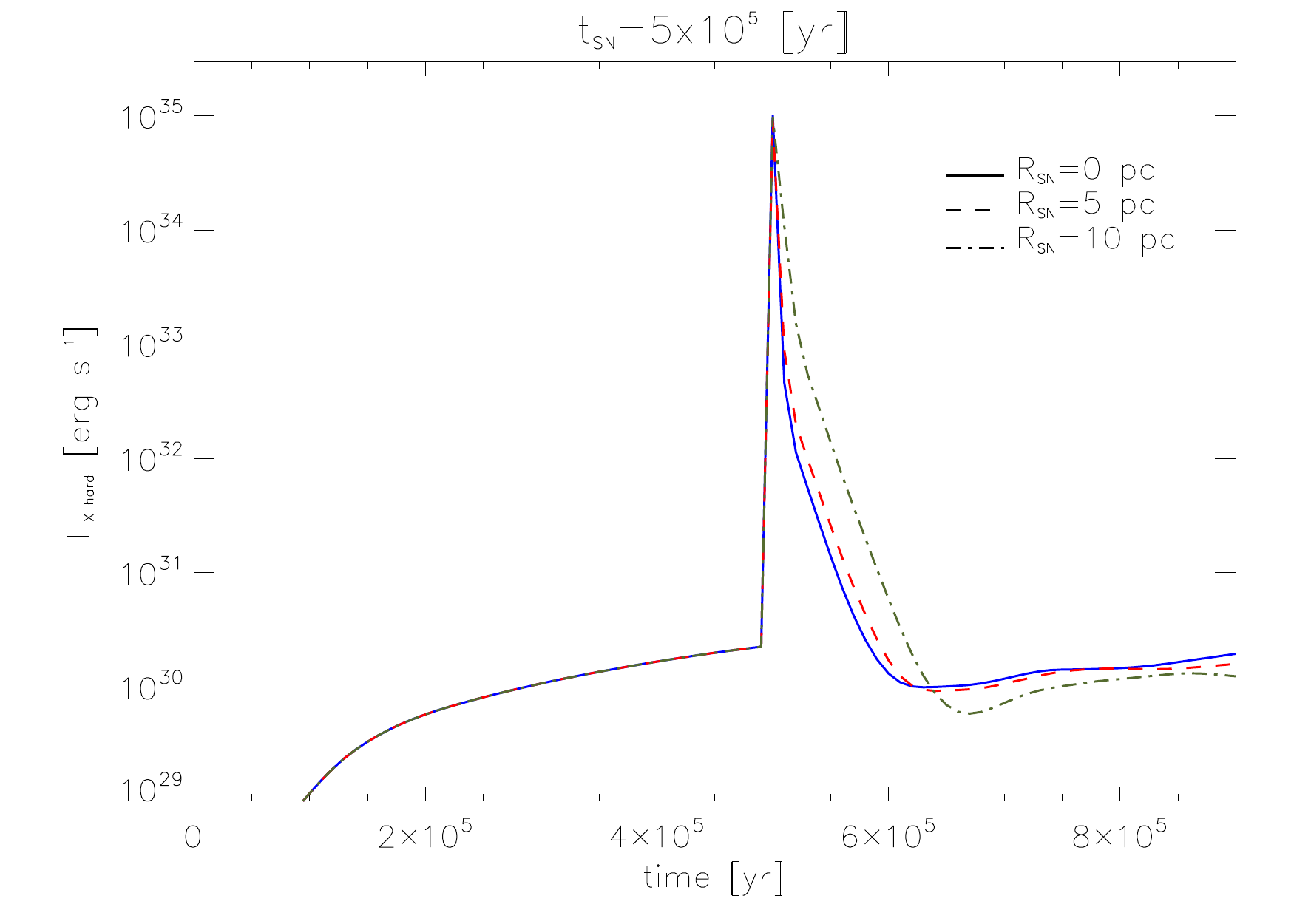}
\caption{Same as Figure~\ref{f:lumXsoftR} for hard X-rays luminosities.}
\label{f:lumXhardR}
\end{figure}

Figure~\ref{f:lumXhardR0C}  shows
the evolution of the hard X-ray luminosity for the two models with
thermal conduction and their counterpart without thermal
conduction. 
The maximum luminosities and the time intervals of high hard X-ray
emission are remarkably similar in spite of the thermal conduction.
Although small differences can be seen, the position of the supernova
explosion in the cluster does not have a significant influence in the
overall and maximum thermal hard X-ray luminosity (see
Figure~\ref{f:lumXhardR}).

It is also intresting to note that for an SN exploding at the centre of the
cluster the highest luminosity achieved is much lower if the explosion
occurs at later times (c.f. the long dashed curve in Figure
\ref{f:lumXhardR0} to the others).

The distribution of the hard X-ray emission before the SN explosion,
at peak luminosity, and after the luminosity returns to its previous
level can be seen as the (logarithmically spaced) contour
levels in Figure \ref{f:Xrays}. The emission during peak luminosity
(middle row) is slightly more extended in the case where the SN is
detonated off-centre (panel f), but returns to being concentrated
inside the star cluster after the effect of the explosion has had time
to decay (bottom row). The impact of thermal conduction on the hard
X-ray emission is also evident. As can be seen, before (top row) or
some time after (bottom row) the SN explosion the hard-band emission
is at a much higher level in the cases without thermal conduction
(left and right columns). However, during the luminosity peak (middle
row) all models display important hard X-ray emission regardless of
the inclusion of thermal conduction. The difference lies mainly in
that in the case including thermal conduction (panel e) the emission
is slightly more centralised than in the cases without. 

We must note that the resolution used in the models is not enough
  to capture all the details of the flow. The use of higher resolution
  allows larger compression factors as well as more small scale structure
  inside the bubbles. To estimate the uncertainty in the X-ray
  luminosities due to poor resolution we have taken a test case (model sr0tsn5z0), and
  reproduce the setup in the {\sc Walicxe-3d} code
  \citep{Toledo2014}. This code has adaptive mesh refinement  (AMR),
  which allows to increase the resolution at a lower computational cost,
  but the thermal conduction is not fully implemented. We ran the test
  case at an equivalent resolution of $512^3$ and $1024^3$ cells, and
  while the details of the flow are different, the integrated X-ray
  luminosities seem to reach convergence. The peak luminosity in the
  higher resolution runs is a factor of $\sim2$ larger than in the
  $256^3$ model sr0tsn5z0. And the times above $10^{34}$, $10^{35}~$, and
  $10^{36}~\mathrm{erg~s^{-1}}$ are larger by a factor of $\sim1.5$. 
  All the results presented above have an uncertainty of this order of
  magnitude due to the limited resolution.

\section{Comparisons with observations}

It is useful to compare our numerical models to the
observations of particular superbubbles. We have thus turned our
attention to four superbubbles located in the Large Magellanic Cloud (LMC):
N70, N185, DEM L50 and DEM L152. These superbubbles have some particular
features (as we will show) that make them compatible with our results.

N70 is a superbubble with a radius of approximately 53 pc. According
to the observations there is a SN located closer to the center than
the edge of the cluster.  
The X-ray luminosity reported by \citet{Reyes2014} is about
2.4($\pm0.4$)$\times 10 ^{35}$~erg s$^{-1}$. We can see that our
numerical results are in good agreement with these observations, in
particular the models with the  SN explosion in the center of the
cluster. 

The case of N185 is quite similar to that of N70. N185 has a spherical
shape with an approximate radius of 43 pc \citep{Oey1996}. The X-ray luminosity
obtained by \citet{Reyes2014} is 2.1 ($\pm0.7$)$\times 10^{35}$~erg s$^{-1}$. Following \citet{Rosado1982}, a possibility to explain the high velocity of this superbubble is that a SN explosion ocurred. From its spherical shape, we conclude that the SN explosion must be located near the centre of the cluster. As in the previous case, our numerical models are in good agreement with the observed X-ray luminosities.

Two other interesting cases are DEM L50 and DEM L152. These are two
superbubbles with very intense X-ray emission. According to observations 
DM L152 has a radius of approximately 50 pc \citep{Jaskot2011}, and DEM L50 has roughly the same radius \citep{Oey1996}. \citet{Jaskot2011} reported an X-ray
luminosity in the range 2.0-4.0$\times 10^{36}$~erg s$^{-1}$ for DEM L50 and an
emission of 5.4-5.7$\times 10^{35}$~erg s$^{-1}$. These two
superbubbles contrast with N70 and N185 in that they contain an
off-centre SN explosion, which is clearly distinguishable in the
observations. Only numerical model sn10tsn5zv predicts a luminosity
comparable to the observed values. 
The luminosities predicted by other models are not high enough to
match with these observed values.

\begin{figure*}
\centering
\includegraphics[width=\textwidth]{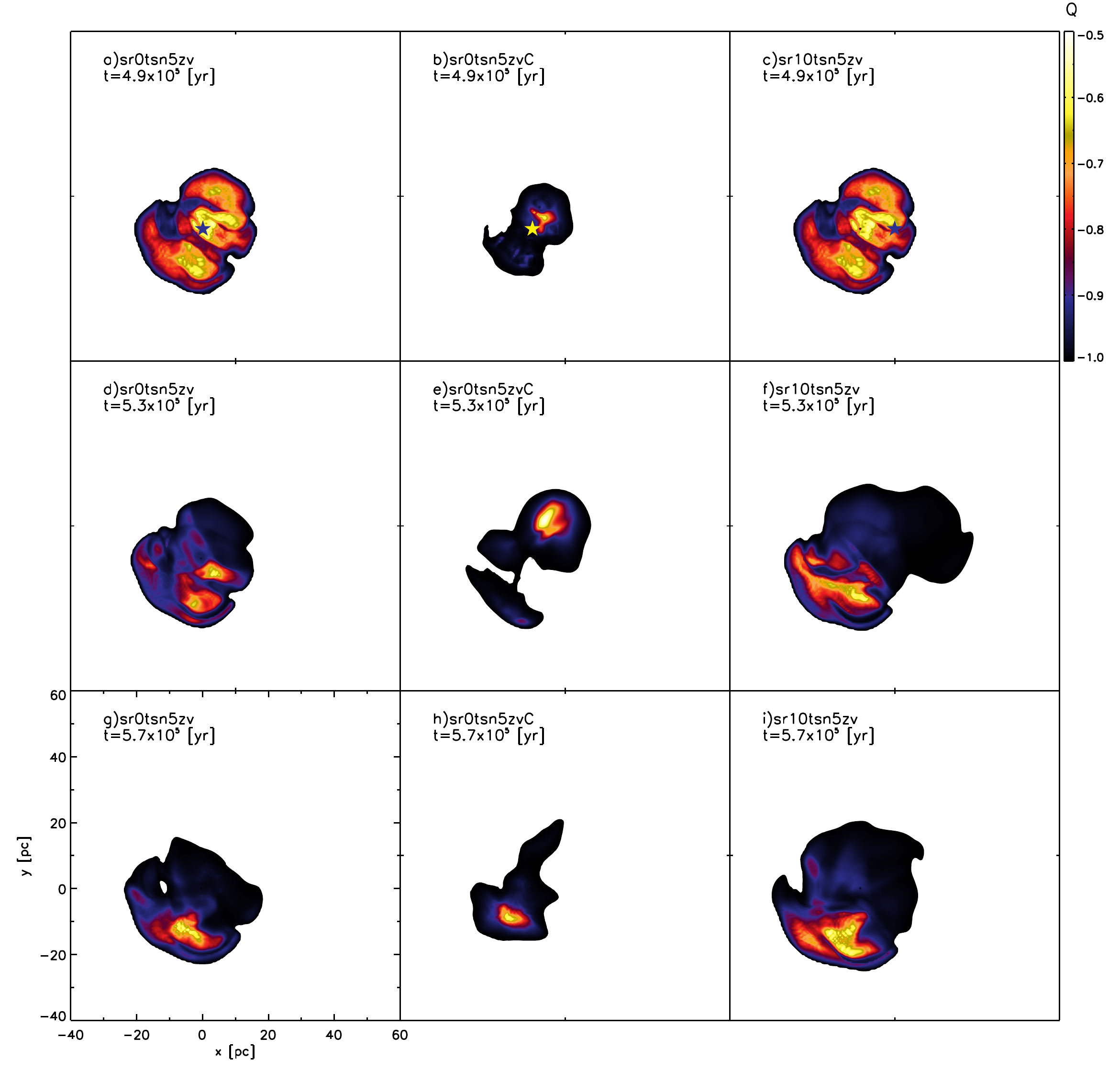}
\caption{Hardness maps for the same cases shown in Figure \ref{f:Den}.}
\label{f:Hardness}
\end{figure*}

In order to better compare with observations, we have calculated the hardness ratio for three of our models (see Figure \ref{f:Hardness}). These models are those discussed in Section 3: two with a centred SN explosion that differ in the presence of thermal conduction, and the third with the SN explosion 10 pc off-centre. Following \citet{Jaskot2011}, the hardness ratio is defined as: 
\begin{equation}
Q=\frac{H-S}{H+S}\, ,
\label{eq:hardness}
\end{equation}
where $H$ is the flux energy in the 2-10~keV energy band (corresponding 
to hard X-rays) and $S$ is the flux energy in the 0.1-2~keV energy band
(corresponding to soft X-rays). After computing the fluxes and obtaining
$Q$ for each cell in the simulation, we integrate along the $z$ axis
in order to project the result into a 2D map, assuming that the X-ray 
absorption due to the material inside the bubble can be neglected.

In Figure \ref{f:Hardness} we observe that, before the SN explosion occurs,
in models without thermal conduction (left and right columns) the hardness
ratio peaks at $\sim-0.55$ at the centre of the stellar distribution, and 
decreases as we approach the edge of the cluster (the shell of swept-up 
ISM material emits mainly soft X-rays). In the case of the model that 
includes thermal conduction (middle column), we observe that $Q\sim -1.$
in most of the bubble, indicating that hard X-ray emission is largely 
negligible. As a result, the material ejected by the stars is cooled
quickly from hard X-ray emitting temperatures ($10^8$ K) to temperatures
in the range $10^5$ - $10^6$ K where soft X-ray emission is favored.

The SN explosion drastically alters the hardness maps for the models 
without thermal conduction. The SN shockwave sweeps the cluster volume, 
devoiding the center of the bubble of hard X-ray emitting gas and 
forming regions with $Q\sim-0.6$ closer to the edge of the bubble. In the 
model with thermal conduction, the hard X-ray emission is small to begin 
with, and the effect of the explosion on the hardness ratio is not as 
noticeable.

The predicted Q values obtained in our numerical models are similar to 
those obtained by \citet{Jaskot2011} for DEM L50 and DEM L152. 
Nevertheless, the specific details and assumptions of our simulations 
make it hard to establish direct comparisons to specific observed 
bubbles. In order to use the hardness ratio to predict some of the physical
processes that occur or have occurred in the super-bubbles we would need to
separately simulate  the specific physical details, such as the position
and mass and energy injection rates of each star, the ISM density, of each
particular bubble, which is out of the scope of this work.

\section{Conclusions}

In this paper we present 3D hydrodynamical models of the evolution
of the soft and hard thermal X-ray luminosities produced inside
superbubbles driven by massive stellar winds including the effect 
of a supernova explosion and thermal conduction.

In all models we include the injection of mass and energy by a
cluster of wind sources and a single SN event. We have varied the
position of the SN with respect to the centre of the cluster as well as
the detonation time. Also, we have worked out models with
a uniform metallicity and models in which
the environment, the winds and the SN have different metallicities.
The metallicities are used to calculate the radiative cooling rate,
and has an effect on the emissivity in X-rays. We have also taken into
account the effects of thermal conduction in two of the models.

In the models with different metallicities
we used $Z=0.3~Z_\odot$ for the environment, $3~Z_\odot$ for the winds
and $10~Z_\odot$ for the SN ejecta.
Our models show that the contribution of the metallicity of the winds and
the supernova remnant is negligible for the soft X-ray emission of 
superbubbles, but becomes important for the hard X-ray component. 
In these models the ratio of soft to hard maximum luminosity can be as
extreme as $10$ (i.e. the  hard X-ray luminosity reaches 10\% of
the one for soft X-rays).

The models with thermal conduction result in  a noticeable
increase in the total luminosity of soft X-rays, by a factor of $\sim
1.25$.
However, this factor is smaller than the two orders of magnitude
difference predicted in the standard
model of \citet{Weaver1977} and \citet{Chu1990}. The differences are likely to
come from the fact that the standard model of \citet{Weaver1977}
considers just a single star, and the extension to a star cluster in
\citet{Chu1990} and \citet{Chu1995} does not account for the cooling of the 
gas due to
the interaction of the stellar winds. Thermal conduction has a slightly larger effect on the total integrated emission of hard X-rays, increasing the luminosity by a factor of $\sim 2.6$.

The most important contribution to the emission of soft and hard
X-rays is produced by the injection of mass and energy by supernova
explosions. In soft X-rays the luminosity increases by up to two orders of
magnitude when we consider a supernova explosion placed at the cluster
centre, and up to three orders when it explodes at the edge the star cluster. 

Another important factor to consider is the time during which the
luminosity remains high (i.e. observable). 
We show that when off-centre supernova events occur
(close to the shell) the luminosity can increase by one or two orders
of magnitude above that predicted by the standard model without SN, 
 and that it can be maintained by a few tens of thousands of
years. Indeed, as the  supernova explosion occurs closer to the shell
of swept up ISM, the maximum luminosity of soft X-rays as well as the
time interval during which luminosity is enhanced increase.

An important increase in the maximum soft X-ray luminosity is produced
when the SN ejecta collide with the dense shell of swept up ISM gas left behind by its interaction with the cluster
wind. In these  cases  X-ray luminosities of
$10^{36}$~erg~s$^{-1}$ can be achieved.
On the other hand, superbubbles where the SN explosions have not
taken place near the shell, such as N\,70 and N\,185
(\citealt{Jansen2001} and \citealt{Reyes2014}), have lower X-ray
luminosity, and can be explained using our models with a slightly off-centre SN. 

In clusters without SN events, or with a SN placed at the centre of
the cluster, the contribution to the luminosity made by the SN is
hard to observe, in particular because the observable flux increase in
the soft X-ray emission lasts for a short time. This could be happening in massive stellar clusters in the Galaxy, such as Arches, Quintuplet and NGC\,3603, 
that have a hundred massive stars with a total observed X-ray emission of 
$\sim 10^{34}$~erg~s$^{-1}$. 

\section*{acknowledgements}

We acknowledge support from CONACyT grants 167611 and 167625, and DGAPA-UNAM
grant IG100214 and IN 109715 . We thank the anonymous referee for very relevant comments that resulted in a substantial revision of the original version of this paper.

\bsp
\label{lastpage}
\end{document}